{
\documentclass[onecolumn,showpacs,amsmath,amssymb,aps]{revtex4}

\usepackage{graphicx}
\usepackage{dcolumn}
\usepackage{bm}

\usepackage{amsmath}
\usepackage{amssymb}
\usepackage{mathrsfs}
\usepackage{url}
\usepackage{subfigure}
\usepackage{algorithm}
\usepackage{algorithmic}
\usepackage{enumerate}
\usepackage{slashbox}

\begin{document}

\preprint{APS/123-QED}

\title{Joint Nonnegative Matrix Factorization for Community Structures Detection in Signed Networks}

\author{Chao Yan}
\affiliation{School of Statistics and Mathematics, Central University of Finance and Economics}

\author{Hui-Min Cheng (co-first author)}%
\affiliation{School of Statistics and Mathematics, Central University of Finance and Economics}

\author{Xin Liu}%
\affiliation{School of Statistics and Mathematics, Central University of Finance and Economics}

\author{Zhong-Yuan Zhang}%
\email{zhyuanzh@gmail.com}
\affiliation{School of Statistics and Mathematics, Central University of Finance and Economics}

\date{\today}

\begin{abstract}
Community structures detection in signed network is very important for understanding not only the topology structures of signed networks, but also the functions of them, such as information diffusion, epidemic spreading, etc. In this paper, we develop a joint nonnegative matrix factorization model to detect community structures. In addition, we propose  modified partition density to evaluate the quality of community structures. We use it to determine the appropriate number of communities. The effectiveness of our approach is demonstrated based on both synthetic and real-world networks.
\end{abstract}

\pacs{Valid PACS appear here}
\maketitle

\section{\label{Introduction}Introduction}
Many complex systems can be modeled as   networks\cite{strogatz2001exploring,watts1998collective}, where the node denotes an individual and the edge represents the relationship between  individuals.
Intuitively, the relationship can be either positive or negative. Take social network for example, people involved in discussing a topic can be connected positively or negatively, depending on whether they agree or disagree with each other.
Such network with  positive and negative edges is called signed network \cite{doreian1996partitioning}.
Positive edge reveals the positive relationship, such as "friendship", "agreement", "trust", yet the negative one reveals the negative relationship, such as "hostility", "disagreement", "distrust". 

Recent years have witnessed increasing interest in community detection of complex networks, which sheds light on how real networks operate\cite{cai2014discrete}.
Community structures are proposed by  Girvan and Newman  for the first time, and refers to groups with dense intra-links and sparse inter-links\cite{girvan2002community,newman2006modularity}.
 However,  community structures in signed network are not only  determined by density but also signs of the links\cite{yang2007community,chen2014overlapping}.
 That is, most intra-links are positive, and  most inter-links are negative at the same time\cite{Harary1953On,yang2007community}.
Due to the fact that it is natural to have some negative links within communities, and some positive links between communities, detecting community structures in signed network poses greater challenges\cite{liu2014multiobjective,yang2007community}.

 Facing the challenges, some efforts have been made to identify the community structures in signed network. Yang et al. proposed an agent-based random walk model to mine the community structures in signed network \cite{yang2007community}. Anchuri et al. generalized spectral approach with iterative optimization to explore the community in signed network \cite{Anchuri2013Communities}. G\'{o}mez et al. firstly generalized the modularity to signed network \cite{gomez2009analysis}, and Traag et al. proposed modularity-optimization based algorithm \cite{traag2009community}.
 Chen et al. proposed a novel approach named as signed probabilistic mixture (SPM) model for overlapping community detection \cite{chen2014overlapping}.
Recently, other statistical inference approaches, such as SBM and SISN, have been generalized to signed network\cite{jiang2015stochastic,zhao2017statistical}.
However, among these previous works, some of them fail to  get accurate partition for large networks with thousands of nodes \cite{yang2007community,Anchuri2013Communities,zhao2017statistical}, while others need predefined community number\cite{chen2014overlapping}.
Thus, we encounter two critical problems: (a) How to detect community structures of signed network accurately, and (b) how to determine the optimal  community number when we have no prior knowledge of community number?

Motivated by  these questions, we develop a novel model, \underline{J}oint \underline{N}onnegative \underline{M}atrix \underline{F}actorization (JNMF for short). Moreover, we propose modified partition density to determine the optimal community number\cite{ahn2009communities,zhang2013overlapping}. In this way, our methods try to solve problems (a) and (b). To evaluate the effectiveness of our approach, we conduct experiments on both synthetic and real-world networks. The experimental results show that our approach is more effective compared with the state-of-the-art approaches.



\section{\label{JNMF}JNMF Method Description}
\subsection{Model Formulation}

Nonnegative Matrix Factorization (NMF) is one of the most popular methods for unsupervised learning \cite{brunet2004metagenes,lee1999learning,lee2001algorithms}, and community structures detection in unsigned network can be naturally formulated as the following tri-factor NMF model: 
\begin{equation*}
\begin{aligned}
   &\min_{W,H}  \sum_{i,j} \left(A_{ij}-(HWH^T)_{ij} \right)^2 \\
  &s.t. \quad H\in \mathbb{R}_+^{n\times c} \quad W\in \mathbb{R}_+^{c\times c},\\
  &\hspace{9mm} \sum_{r=1}^c H_{ir}=1,
\end{aligned}
\end{equation*}
where $A$ is the adjacency matrix of size $n\times n$,
$H$ is the community membership matrix of size $n\times c$ where element $H_{ir}$ is the probability of node $i$ in community $r$,
$W$ is the community-relation matrix of size $c\times c$ where element $W_{rs}$ is the probability of edges existing between communities $r$ and $s$, $n$ is the number of nodes in network, and $c$ is the community number.

In this paper, we extend the above model to joint tri-factor nonnegative matrix factorization model (JNMF) to detect communities in signed networks, which are (a) intra-connected mainly by positive edges and (b) inter-connected mainly by negative ones.

Given a signed undirected network $\mathscr{G}$, its adjacency matrix $A$ is a symmetric  matrix with mixed signs representing different relations among the nodes.
$A$ can be separated into two matrices $A^+$ and $A^-$ containing the patterns (a) and (b), respectively, and $A= A^+ - A^-$, where
 \begin{equation*}
 A_{ij}^+= \left\{
   \begin{aligned}
    A_{ij},&\quad \text{if}~ A_{ij}>0\\
     0,&\quad \text{otherwise}\\
    \end{aligned}
  \right.
 \end{equation*}

\begin{equation*}
 A_{ij}^-= \left\{
   \begin{aligned}
    -A_{ij},&\quad \text{if}~ A_{ij}<0\\
     0,&\quad \text{otherwise}\\
    \end{aligned}
  \right.
\end{equation*}

\begin{figure}[h]
\hspace{7mm}
 \includegraphics[width=15cm,height=8cm]{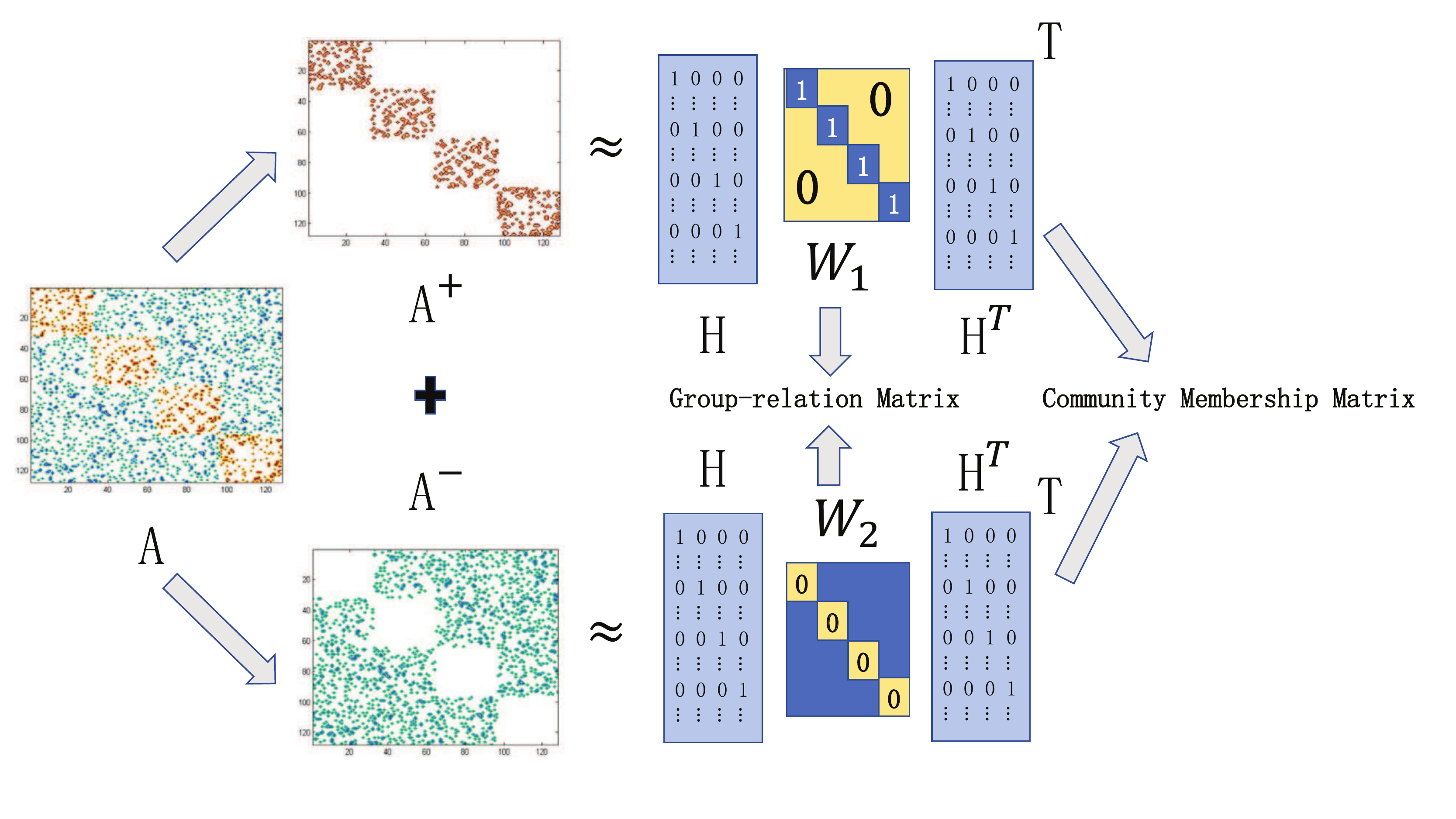}
\caption{ The flowchart illustrates how JNMF works to figure out community structures in signed networks.}\label{motivationFormulation}
\end{figure}

%

%

By discovering the hidden patterns behind $A^+$ and $A^-$, JNMF can detect community structures in signed networks:
\begin{equation}\label{eq:1}
\begin{aligned}[rc]
  \hspace{3mm}\min_{H,W} & \hspace{3mm} \left\{\frac{1}{2} \sum_{i,j} \left[(A^{+}_{ij}-(HW_{1}H^T)_{ij} ) \right]^2+\right. \\\vspace*{4mm}
  & \hspace{15mm}\left.\frac{1}{2} \sum_{i,j}  \left[(A^{-}_{ij}-(HW_{2}H^T)_{ij} ) \right]^2\right\} \\
  s.t. & \hspace{3mm} H\in \mathbb{R}_+^{n\times c}, \quad W_1\in \mathbb{R}_+^{c\times c}, \quad W_2\in \mathbb{R}_+^{c\times c},\\
   &\hspace{3mm}\sum_{r=1}^c H_{ir}=1, \\
   & \hspace{3mm} W_1 \mbox{ is diagonal}, \\
   &\hspace{3mm} W_2 \mbox{ is with all diagonal entries zero},
   \end{aligned}
\end{equation}
where $W_1$ and $W_2$ are the $c\times c$ community-relation matrix of $A^+$ and $A^-$ respectively.

Noting that there are three kinds of entries in $A$: a) $A_{ij} > 0$ means that there is positive relation between the two nodes; b) $A_{ij}<0$ means that there is negative relation between the two nodes; and c) $A_{ij}=0$ means that there is no relation between the two nodes or no information on the relation between them. Obviously, the first two kinds of entries have higher priorities than the third one for community structures detection. Hence we introduce weight matrix $B$ into the model (\ref{eq:1}) for better numerical results:
\begin{equation}\label{eq:1.1}
\begin{aligned}[rc]
   \min_{H,W} & \hspace{3mm} \{ \frac{1}{2} \sum_{i,j} [B_{ij} \circ (A^{+}_{ij}-(HW_{1}H^T)_{ij} ) ]^2+\\
    & \hspace{10mm}+\frac{1}{2} \sum_{i,j}  [B_{ij} \circ (A^{-}_{ij}-(HW_{2}H^T)_{ij} ) ]^2\} \\
  s.t. & \hspace{3mm} H\in \mathbb{R}_+^{n\times c}, \quad W_1\in \mathbb{R}_+^{c\times c}, \quad W_2\in \mathbb{R}_+^{c\times c},\\
   &\hspace{3mm}\sum_{r=1}^c H_{ir}=1, \\
   & \hspace{3mm} W_1 \mbox{ is diagonal}, \\
   &\hspace{3mm} W_2 \mbox{ is with all diagonal entries zero},
   \end{aligned}
\end{equation}
where
\begin{equation*}\label{eq:2}
 B_{ij}= \left\{
   \begin{aligned}
   5,&\quad A_{ij}\neq0\\
     1,&\quad otherwise,\\
    \end{aligned}
  \right.
  \end{equation*}
and is of size $n\times n$.
``$\circ$'' is element-wise multiplication.

\subsection{Algorithm description}
We design multiplicative update rules to solve (\ref{eq:1.1}), which is gradient descent based and is summarized in Algorithm \ref{alg:Framwork}.
\begin{algorithm}[H]
\caption{Multiplicative update rules for JNMF}
\label{alg:Framwork}
\begin{algorithmic}[1] 
\REQUIRE ~~\\ 
Adjacency matrix $A$ of signed network $\mathscr{G}$;\\\vspace{2mm}
Weighting matrix, $B$;\\\vspace{2mm}
Iteration number, iter;\\\vspace{2mm}
Community number, $c$;\\\vspace{2mm}
\ENSURE ~~\\\vspace{2mm} 
Community membership matrix $H$;\\\vspace{2mm}

\STATE \textbf{for} $t = 1 : iter$ \textbf{do}\\\vspace{2mm}
\STATE $\displaystyle W_1:=(W_1)\frac{(H^T (B\circ A^{+})H)}{(H^T(B\circ HW_1 H^T)H)}$\\\vspace{2mm}
\STATE $\displaystyle H:= H \frac{((B\circ A^{+})HW_1^T)+((B\circ A^{-})HW_2^T)}{(B\circ HW_1 H^T) HW_1^T+(B \circ HW_2 H^T )HW_2^T} $\\\vspace{2mm}
\STATE $\displaystyle H_{ir} := \frac{H_{ir}}{\sum_{r=1}^c H_{ir}}$\\\vspace{2mm}
\STATE $\displaystyle W_2:=(W_2) \frac{(H^T (B\circ A^{+})H)}{(H^T(B\circ HW_1 H^T)H)}$\\\vspace{2mm}
\STATE $\displaystyle H:= H \frac{((B\circ A^{+})HW_1^T)_{ir}+((B\circ A^{-})HW_2^T)}{(B\circ HW_1 H^T) HW_1^T+(B \circ HW_2 H^T )HW_2^T} $\\\vspace{2mm}
\STATE $\displaystyle H_{ir} := \frac{H_{ir}}{\sum_{r=1}^c H_{ir}}$\\\vspace{2mm}
\label{code:fram:select}
\RETURN ${H}$;\\\vspace{1mm} 
\end{algorithmic}
\end{algorithm}

\section{\label{selection}Model Selection}

For real-world signed network, we have no information of community number, posing an impediment to real application.
To address this problem, several criteria, such as modularity $Q$\cite{newman2004finding,newman2006modularity} and partition density\cite{ahn2009communities,zhang2013overlapping}, have been proposed to evaluate the quality of the detected communities and to choose appropriate community number, where the criterion achieves its maximum.

In this section, we extend the partition density for signed networks. Given a signed network, the partition density $D_{\alpha}$ of community $\alpha$ is defined as:
\begin{equation*}
D_{\alpha}= \frac{m^+_{\alpha}-m^-_{\alpha}}{n_{\alpha}(n_{\alpha}-1)/2},
\end{equation*}
where $m^+_{\alpha}$  is the number of positive links within community $\alpha$, and $m^-_{\alpha}$ is the number of negative ones,
$n_\alpha$ is the number of nodes in community $\alpha$, and the overall partition density of the signed network is the average of $D_{\alpha}, \alpha = 1, 2,...., c$, weighted by the  the fraction of nodes in each community:
\begin{equation*}
D =\displaystyle{\sum_{\alpha=1}^c \frac{n_\alpha}{N}D_\alpha},
\end{equation*}
Partition density has inverse resolution limit problem, and preference towards small communities\cite{lee2017inverse}, hence we add a penalty term into the denominator to control this problem, and the definition is updated as follows:
\begin{equation}\label{eq:pd4}
D =\displaystyle{\frac{\displaystyle\sum_{\alpha=1}^c \frac{n_\alpha}{N}D_\alpha}{\sqrt{c}}}.
\end{equation}

\section{\label{example}An Illustrative Example}

\begin{figure}[h]
\centering 
\includegraphics[width=15cm,height=8cm]{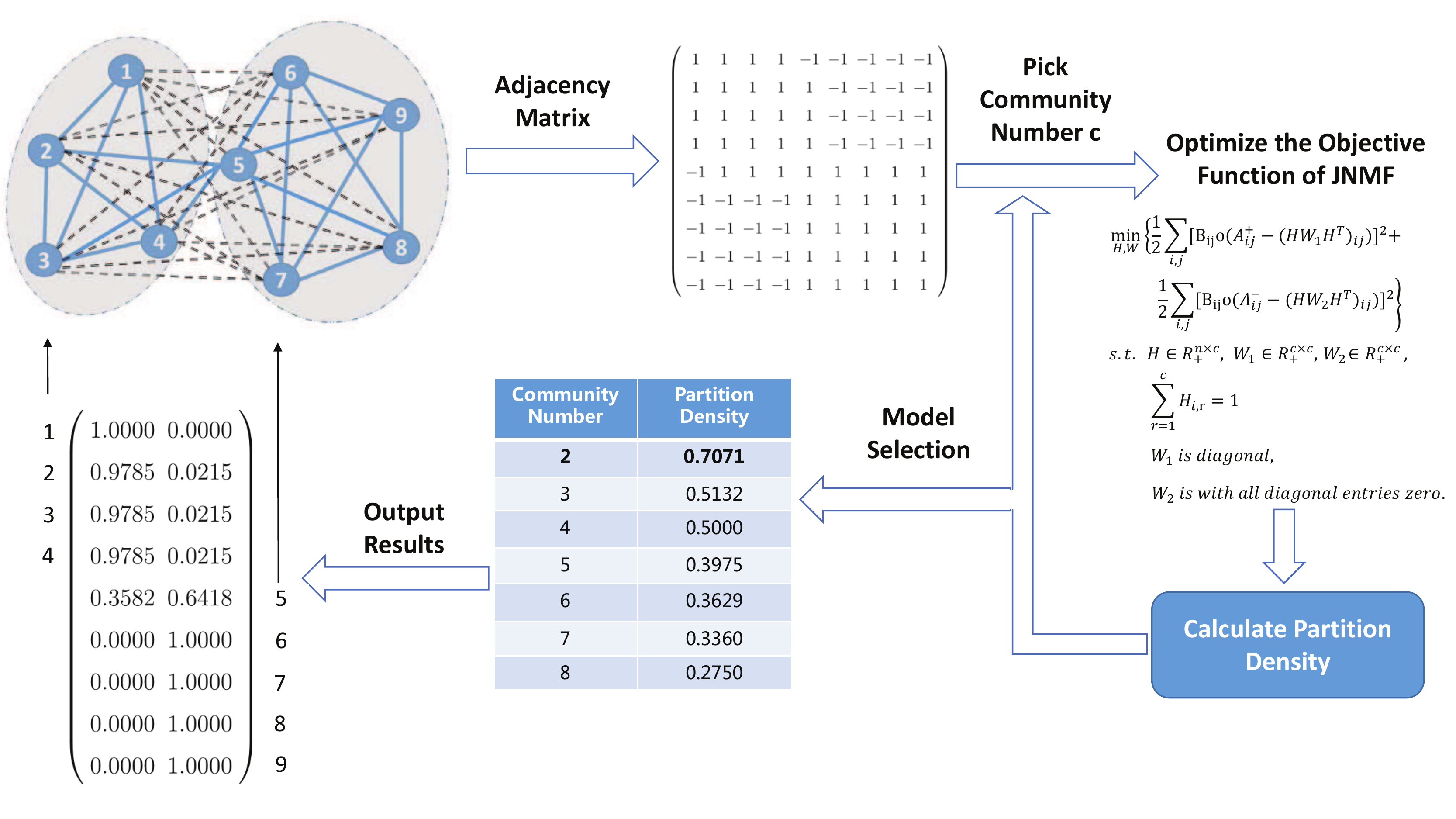}
\caption{A small example demonstrating how the proposed method works.
Solid lines in the network are positive (supportive) relations, and dashed ones are negative (opposed) relations.}\label{illustrative}
\end{figure}

Fig.\ref{illustrative} uses the U.S. supreme court justices network with two communities\cite{doreian2009partitioning,su2017algorithm} to demonstrate how the proposed method works. This network presents the voting behavior of nine justices in the U.S. supreme court between 2006 and 2007, where the positive links mean agreement when voting and the negative ones mean disagreement.
We solve the JNMF model (\ref{eq:1.1}) with different community numbers and select the optimal number where the partition density achieves its maximal value. Then the corresponding matrix $H$ is output as the final result, and node $i$ is assigned into community $r$ if $H_{ir}$ is the maximal value of the $i$th row.

Finally, we get two communities, $\{1, 2, 3, 4\}$ and $\{5, 6, 7, 8, 9\}$, which agree with  the background information of the network.

\section{\label{experiments}Experimental Results}
In this section, we use both the synthetic and real world networks to demonstrate the effectiveness of the proposed method.
\subsection{Datasets Description}

\begin{enumerate}
\item SG benchmark network \cite{yang2007community}: The SG benchmark network grows out of Girvan-Newman network (GN network), and has six parameters $c,n,k,p_{in},p_+,p_-$. $c$ is the community number, $n$ is the number of nodes in each community, $k$ is the average degree  in the network, $p_{in}$ denotes the
probability of internal links, and $p_+, p_-$ denote the fraction of positive inter-links and negative intra-links respectively, which are also named as noise level. The community structures become less clear and more difficult to be detected with decreasing $p_{in}$  and increasing noise level. In this paper, we set the parameters as follows: $c = 4, n = 30, k =16$ and generate two kinds of SG networks:

(1) SG\_1: There is no noise in SG\_1, i.e., $p_+ = 0$ and $p_- = 0$. The parameter $p_{in}$ is from 0 to 1.

(2) SG\_2:  Noise is added with different levels, i.e., $p_+\in[0, 0.5]$ and $p_-\in[0, 0.5]$. The parameter $p_{in}$ is set to 0.8.

\item SLFR benchmark network: To address the characteristics of real world networks, the signed Lancichinetti-Fortunato-Radicchi (LFR) benchmark network\cite{lancichinetti2008benchmark,esmailian2015community} is proposed and has ten parameters, $n,k_{avg},k_{max},\lambda_{1},\lambda_{2},s_{min},s_{max},\mu,p_+,p_-$, where $n$ is the number of nodes, $k_{avg}$ and $k_{max}$ represent average degree and max degree respectively, $\lambda_1$ and  $\lambda_2$ mean exponent of power-law distributions of nodes degree and community size respectively, $s_{min}$ and $s_{max}$ mean the minimum and maximum of community number respectively, $\mu$ is the fraction of edges connecting the neighbors in other communities, $p_+$ and $p_-$  are noise level. In this paper, we set the parameters as follows: $n = 1000, k_{avg} = 20, k_{max} = 50, \lambda_1 = 2, \lambda_2 = 1, s_{min} = 20, s_{max} = 60$, and generate two kinds of SLFR networks:

(1) SLFR\_1: There is no noise in SLFR\_1, i.e., $p_+ = 0$ and $p_- = 0$. The parameter $\mu$ is from 0.1 to 0.9.

(2) SLFR\_2: Noise is added with different levels, i.e., $p_+\in[0, 0.5]$ and $p_-\in[0, 0.5]$. The parameter $\mu$ is set to 0.2.

\item Slovene parliamentary party network \cite{ferligoj1996analysis}: The network is about the relations among the ten parties in Slovene parliament, 1994 \cite{ferligoj1996analysis}. It has two communities: $(1, 3, 6, 8, 9)$ and $(2, 4, 5, 7, 10)$\cite{wu2012examining}. The weights of the links in the network were estimated by experts on parliament activities meaning the relations among the parties.
\item Gahuku-Gama subtribes network \cite{read1954cultures}: The network is about the culture of New Guinea Highland\cite{read1954cultures}.
 There are 16 subtribes in this network falling into three communities: $(3,4,6,7,8,11,12 )$, $(1,2,15,16)$ and $(5,9,10,13,14)$ \cite{doreian1996partitioning}, and the positive and negative edges represent political alliance and enmities respectively.
\end{enumerate}

\subsection{Assessment Standard}
We use normalized mutual information (NMI) \cite{strehl2002cluster} to evaluate the performance of our method on synthetic networks, which is defined as follows:
\begin{equation}\label{expected}
I(M_{1},M_{2})=\frac{\displaystyle\sum_{i=1}^k\sum_{j=1}^k n_{ij}log\frac{n_{ij}n}{n_{i}^{(1)}n_{j}^{(2)}}}{\sqrt{\left(
\sum_{i=1}^k n_{i}^{(1)}log\frac{n_{i}^{(1)}}{n}\right)
\left(\sum_{j=1}^k n_{j}^{(2)}log\frac{n_{j}^{(2)}}{n}
\right)
}},
\end{equation}
where $M_{1}$ and $M_{2}$   denote ground-truth and detected community partition respectively, $n_{i}^{(1)}$ and $n_{j}^{(2)}$ are the community size of ground-truth community $i$ and detected community $j$ respectively, and $n$ is the number of all nodes. While $n_{ij}$ counts the number of nodes assigned to detected community $j$, which belong to ground-truth community $i$. The larger the NMI value, the better the detection performance.


\subsection{Experimental Results on synthetic networks:}
In this section, we compare our methods JNMF\_1 and JNMF\_2 with three state-of-the-art methods including FEC, SPM and SISN on synthetic networks. JNMF\_1 is JNMF with community number $c$ predefined to be the ground-truth, and JNMF\_2 is JNMF with $c$ determined by partition density. FEC is an agent-based heuristic method for community detection which does not need a predefined community number \cite{yang2007community}. SPM is the abbreviation for signed probabilistic mixture model and needs a predefined community number\cite{chen2014overlapping}. SISN is a recently proposed statistical inference method for signed networks and does not need a predefined community number \cite{zhao2017statistical}. The results are averages of ten trials. Note that given a network with $n$ nodes, $l^+$ positive links, $l^-$ negative links and $c$ communities, the time complexities of SPM and SISN are
$O(T(l^+\times c + l^-\times c^2))$ and $O(n^4)$ respectively, where $T$ is the iteration steps of EM algorithm\cite{chen2014overlapping,zhao2017statistical}, making them very slow for large scale networks, hence we only output the results of FEC on SLFR benchmark networks. We also compare the abilities of our method with FEC and SISN to infer the appropriate number of communities.

\begin{figure*}
\centering
\includegraphics[width=17cm,height=7cm]{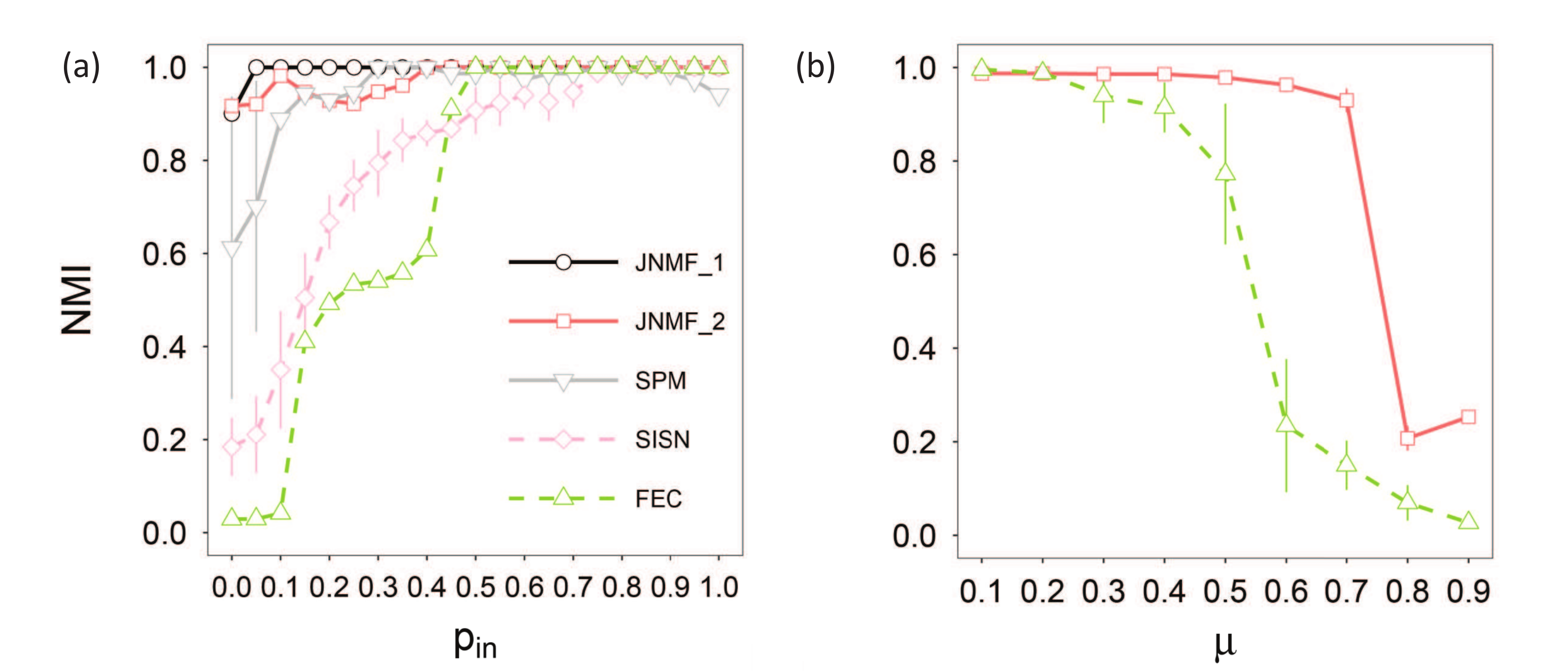}
\caption{Averaged NMI with the standard deviation of different methods on SG\_1 benchmark networks (a) and SLFR\_1 benchmark networks (b). Note that for SPM and FEC in (a), we predefine $c$ to be the ground truth.
}\label{SGN_SLFR_2dim}
\end{figure*}

\begin{figure*}
\centering
\includegraphics[width=17cm,height=7cm]{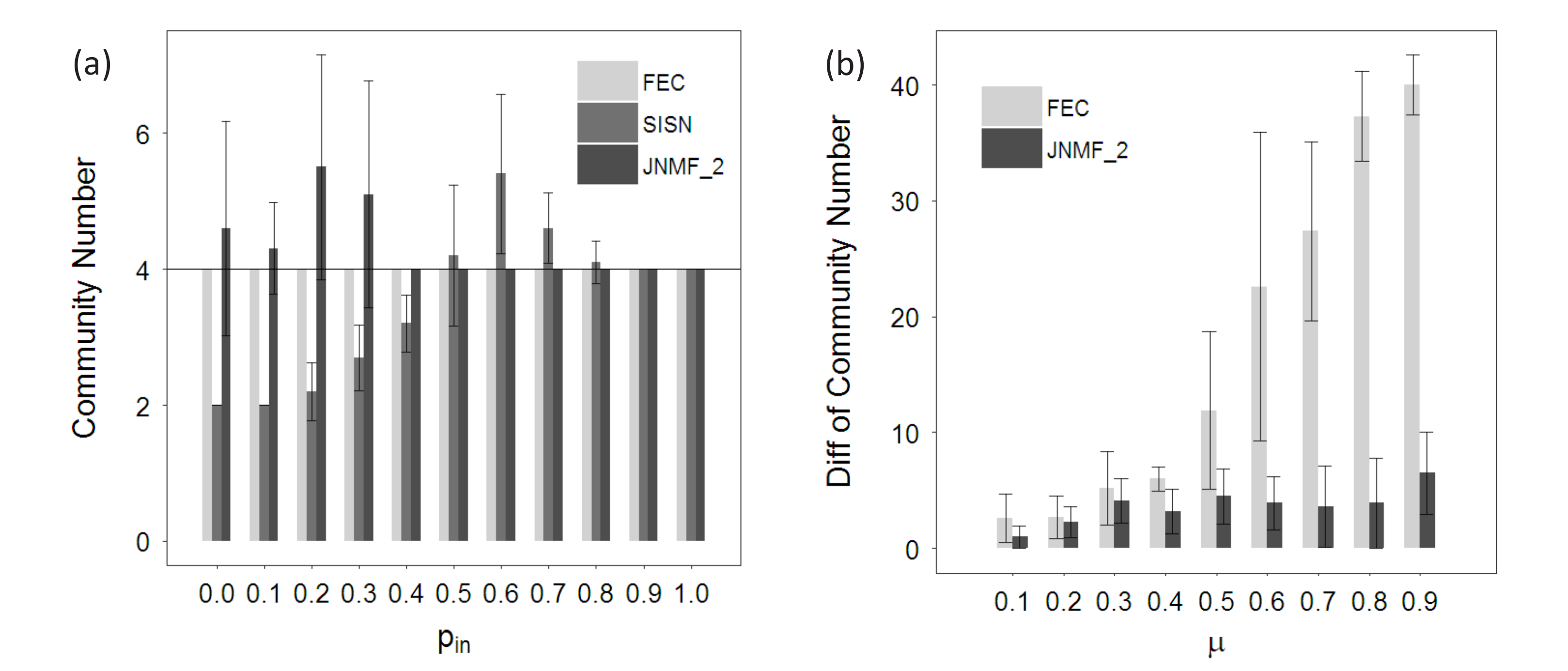}
\caption{ (a) Averaged detected community number with the standard deviation on SG\_1 benchmark networks, and the black horizontal line indicates the ground truth. (b) Averaged difference between detected community number and the ground truth on SLFR\_1 benchmark networks.
}\label{SGN_SLFR_2dim2}
\end{figure*}

Firstly, the experiments conducted on SG\_1 and SLFR\_1 are shown in Fig.\ref{SGN_SLFR_2dim} and Fig.\ref{SGN_SLFR_2dim2}, from which one can conclude that: (i) The NMI results of JNMF is larger than $90\%$, and outperform the other methods in most cases, especially on SLFR\_1. For example, when $\mu$ = 0.6, the NMI of our method (96.28\%) is 73\% higher than that of the FEC (23.44\%). The performance of SPM is slightly better than JNMF\_2 in some cases because SPM uses the true number of communities as input.
(ii) The standard deviations of the proposed methods are lower (close to zero), meaning that JNMF is more stable.
(iii) The inferred community number of JNMF\_2 are closer to the real ones, and the standard deviations are lower.

\begin{figure*}
\centering
\includegraphics[width=15cm,height=7cm]{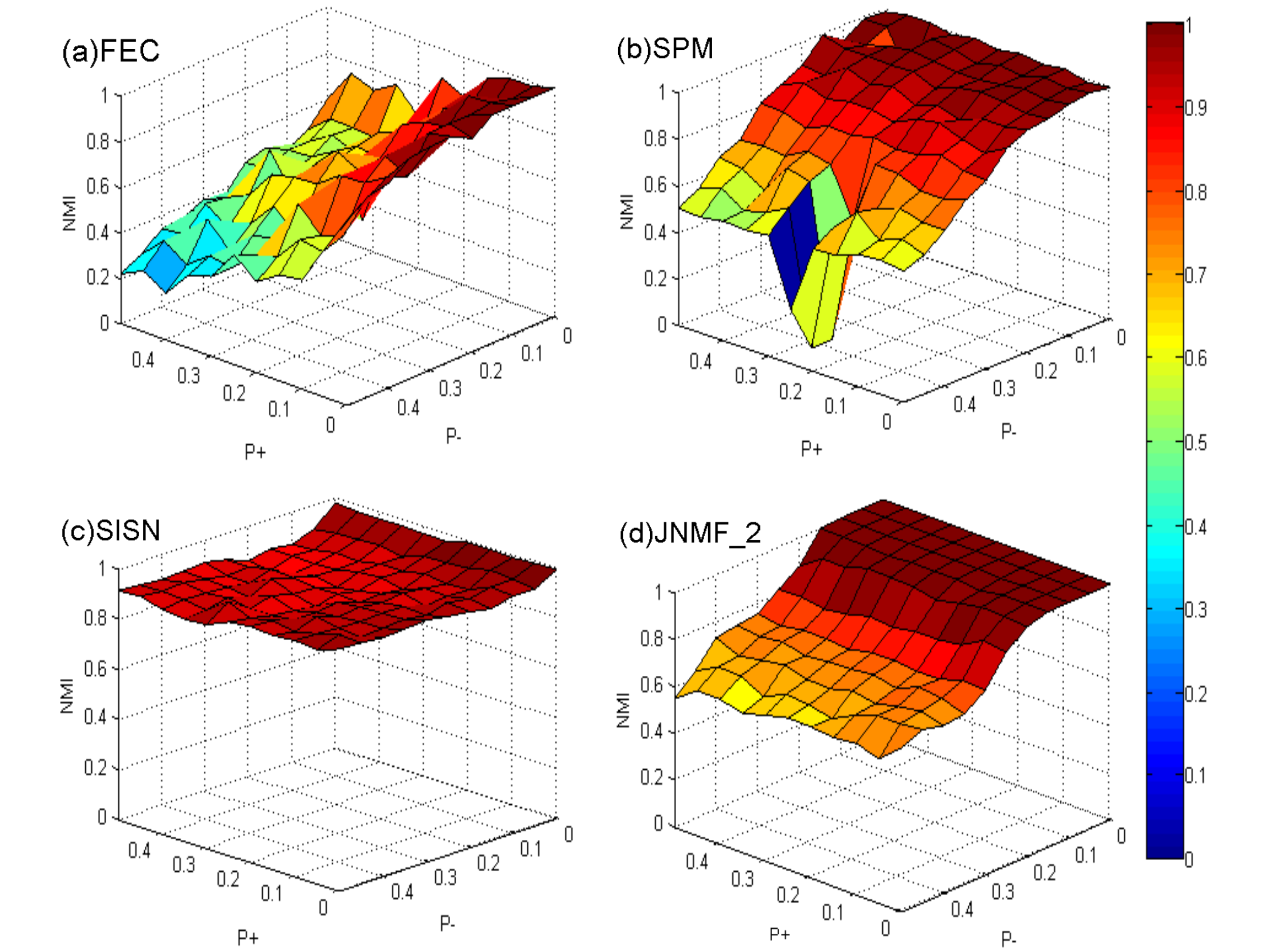}
\caption{ Community detection performance  on SG\_2 benchmark networks of (a) FEC, (b) SPM with predefined c to be the ground truth, (c) SISN and (d) JNMF\_2.
}\label{SGN3dim}
\end{figure*}

\begin{figure*}
\centering 
\includegraphics[width=15cm,height=7cm]{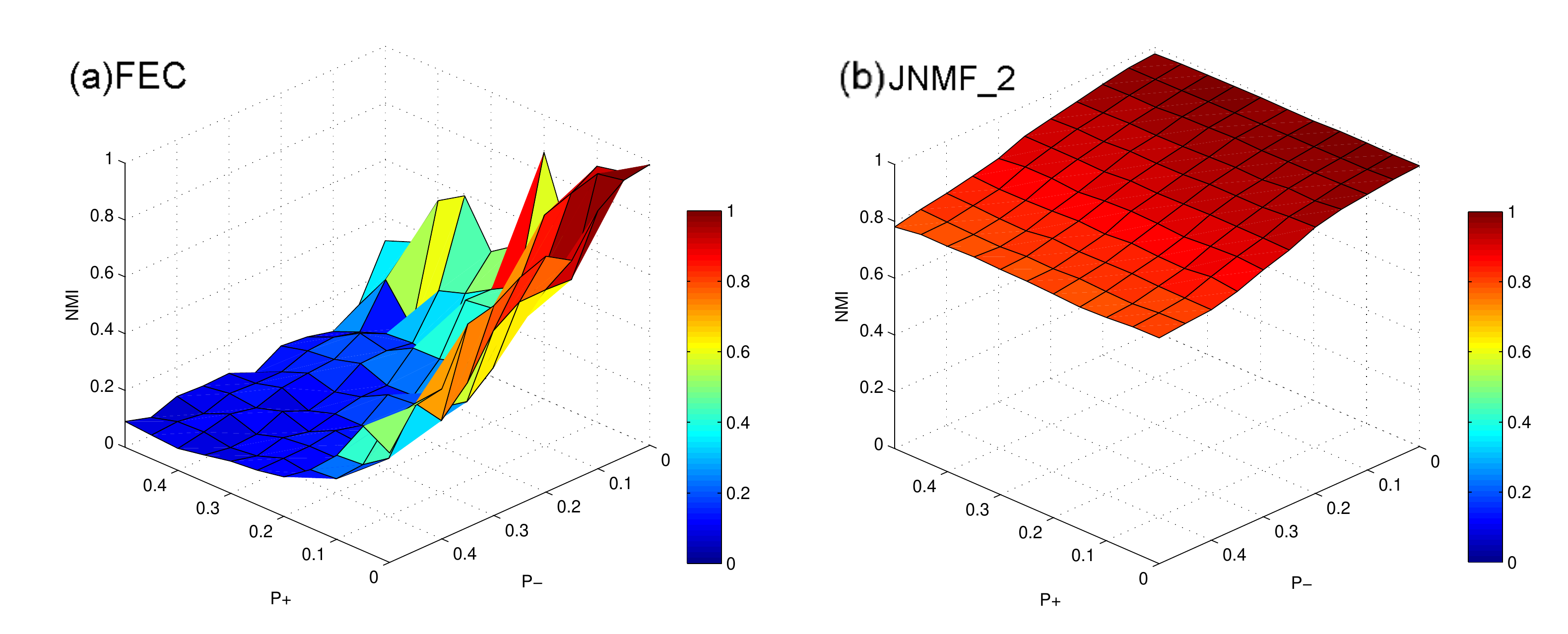}
\caption{Community detection performance  on SLFR\_2 benchmark networks of (a) FEC and (b) JNMF\_2.
}\label{SLFR3dim}
\end{figure*}

Secondly, the experiments conducted on SG\_2 and SLFR\_2 are shown in Fig.\ref{SGN3dim} and Fig.\ref{SLFR3dim}, from which one can conclude that:   (1) JNMF, SPM and SISN are less sensitive to $p_+$ than to $p_-$. However, FEC is sensitive to both $p_-$ and $p_+$.
(2) JNMF is competitive, especially on SLFR\_2 benchmark network, which are more practical.

\subsection{Experimental Results on Signed Real-world Networks}

\begin{figure*}
\centering 
\includegraphics[width=15cm,height=8cm]{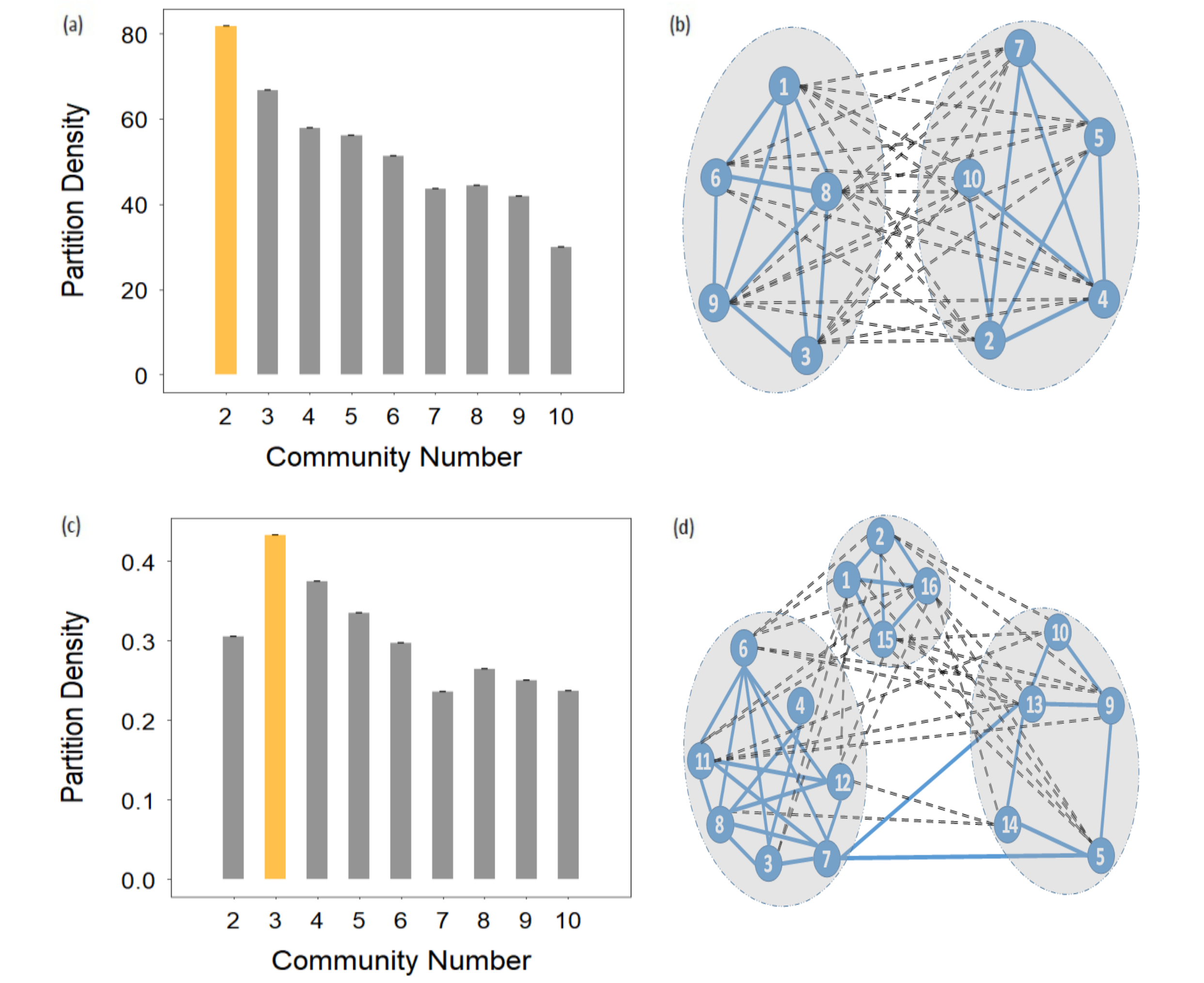}
\caption{
Averaged partition density value with the standard deviation on (a) Slovene parliamentary party network and (c) Gahuku-Gama subtribes network.
 The optimal community number is highlighted in yellow. Community structures detected by our methods on (b) Slovene parliamentary party network and (d) Gahuku-Gama subtribes network. Nodes with the same shape belong to the same group, and  the solid links are positive while the dashed ones are negative.
}\label{SL_GG}
\end{figure*}

To further evaluate the effectiveness of our proposed method, we conduct experiments on real-world networks.
Table \ref{SL_GGtab} gives the
estimated community number of different methods, and Fig.\ref{SL_GG} gives the community structures detection results, from which one can observe that: (i) The inferred numbers of our method are more
reasonable. (ii) The detected communities are identical with domain knowledge and are easy to explain.

\begin{table}
\caption{ Number of communities estimated by different methods.
The last row is the computational time complexity.}
\centering\scriptsize
\begin{tabular}{c c c c c }\hline\hline
\hspace{1mm} $ $ \hspace{1mm} &
 \hspace{1mm} $JNMF$ \hspace{1mm} &
 \hspace{1mm} $SISN$ \hspace{1mm} &
 \hspace{1mm} $FEC$ \hspace{1mm} \\\hline
   Slovene parliamentary party  & 2 $\pm$ 0  & 8.8 $\pm$ 0.42  &  2 $\pm$ 0   \\
   Gahuku-Gama subtribes  & 3 $\pm$ 0   & 3 $\pm$ 0 &  4 $\pm$ 0  \\
   U.S. supreme court   & 2 $\pm$ 0   & 2 $\pm$ 0 &  2 $\pm$ 0  \\
   Time complexity  &  $O(cn^2)$  & $O(n^4)$ &  $O(n^3)$   \\
\hline\hline
\end{tabular}\label{SL_GGtab}
\end{table}

\section{\label{conclusion}Conclusions and Future works}
In this paper, we present a joint nonnegative matrix factorization model to detect community structures in signed networks, and also propose a revised partition density to evaluate the quality of detected communities and to automatically infer the community numbers. The experiments conducted on both synthetic and real-world networks show the effectiveness of the proposed method. In summary, the method is parameter-free, easy to implement.
Interesting problems for future work include generalization of the proposed method to overlapping community detection, multi-view community detection in signed networks.

\bibliography{reference}
\end{document}